\documentclass[printer]{aa}
\usepackage{amssymb}
\usepackage{amsmath}
\usepackage{txfonts}
\usepackage{graphicx}
\usepackage{natbib}
\usepackage{rotating}
\usepackage{soul}
\usepackage{url}
\usepackage{epsfig}
\usepackage{longtable}

\usepackage{color}

\def\gr{$\gamma$-ray}
\def\ee{$e^+e^-$}

\begin{document}

\title{An exploration of hadronic interactions in blazars using IceCube}
\titlerunning{An exploration of the impact of IceCube}

\author{ C. Tchernin$^1$,  J.A. Aguilar$^2$, A. Neronov$^1$, T. Montaruli$^2$}
\institute{ 1. D\'epartement d'astronomie, Universit\'e de Gen\`eve, CH-1290 Versoix, Switzerland.\\
2. D\'epartement de physique nucl\'eaire et corpusculaire,
Universit\'e de Gen\`eve, CH-1211 Gen\`eve 4, Switzerland.}

\keywords{Active Galactic Nuclei, blazars,  Cosmic neutrinos, hadrons.}

\abstract
{Hadronic models, involving matter (proton or nuclei) acceleration in blazar jets, imply high energy photon and neutrino emissions due to interactions of high-energy protons with matter and/or radiation in the source environment.}
{This paper shows that the sensitivity of the IceCube neutrino telescope in its 40-string configuration (IC-40) is already at the level of constraining the parameter space of purely hadronic scenarios of activity of blazars.}
{Assuming that the entire source power originates from hadronic interactions, and assuming that the model describe the data, we estimate the expected neutrino flux from blazars based on the observed $\gamma$-ray flux by Fermi, simultaneously with IC-40 observations. We consider two cases separately to keep the number of constrainable parameters at an acceptable level: proton-proton or proton-gamma interactions are dominant. Comparing the IC-40 sensitivity to the neutrino flux expected from some of the brightest blazars, we constrain model parameters characterizing the parent high-energy proton spectrum.}
{Using together the Fermi photon observations and the IC-40 neutrino sensitivities, we find that when $pp$ interactions dominate, some constraints on the primary proton spectrum can be imposed. For instance, for the tightest constrained source 3C 454.3, the very high energy part of the spectra of blazars is constrained to be harder than $E^{-2}$ with cut-off energies in the range of $E_{cut}\ge 10^{18}$~eV.   When interactions of high-energy protons on soft photon fields dominate, we can find similarly tight constraints on the proton spectrum parameters if the soft photon field is in the UV range or at higher energy, e.g. if it originates from the ``big blue bump'' produced by the accretion disk. The tightest constraint is for 3C 273, with the cut-off energy constrained to be $E_{cut} \gtrsim 10^{18}$~eV for any spectral index and different soft photon fields, including the radiation from the accretions disk, broad line region  or the synchrotron radiation from the jet. } 
{We have adopted a simplified approach using the IC-40 sensitivity (about one half of the full detector) to constrain parameters of blazar hadronic models. Data of the full detector are already available and corresponding constraints should be about a factor of 3 better than what discussed here.
}
\maketitle

\section{Introduction}

Blazars are a special type of radio-loud Active Galactic Nuclei (AGN) emitting jets aligned along the line of sight \citep{urry}. Observations of broad-band electromagnetic emission from blazars from the radio to the TeV region show that the jets and/or central engines of blazars and, more generally, radio-loud AGNs, 
accelerate electrons to extremely high energies. The origin of these electrons and their production sites are not completely understood. The most common considered possibility is that electrons are accelerated at the shocks propagating through the parsec-scale jets. However, observations of fast variability of the highest energy emission indicate that the acceleration site might be located close to the blazar central engine, the supermassive black hole \citep{celotti98,m87,mrk501,pks2155,3c273_fermi,4C2135,4C2135magic}. 

Protons should be accelerated together with electrons, presumably with higher efficiency due to less important energy losses. However, a direct test of the presence of high-energy protons in the source is difficult because of the absence of a clear signature of proton-generated emission in the electromagnetic component of blazar and radio galaxy spectra. The main energy loss channels for the high-energy protons in the AGN environment are pion production in interactions with the low-energy protons \citep{nellen93,bednarek97,beall99,sibiryakov08,neronov09,barkov10}  or interactions with radiation fields \citep{mannheim92,mannheim93,bednarek99,proton_synchrotron,atoyan01,atoyan03,neronov_semikoz}. An additional possibility is proton synchrotron radiation energy losses \citep{proton_synchrotron,aharonian00,mucke03}. Pion production reactions result in emission of \gr s in the decays of neutral pions and in production of neutrinos and electrons/positrons in the decays of charged pions. 
The pion decay \gr\ emission is directly observable by telescopes on the Earth up to the TeV energies. 

A limited possibility to test the hadronic nature of high-energy activity of AGN might occur if the  Ultra-High-Energy Cosmic Rays (UHECR) observed at the Earth are produced by particular nearby radio galaxies. In this case a constraint on the overall energy balance of the sources, including the power in accelerated protons, could be inferred from the observed UHECR flux. In such a situation, predictions for the flux of neutrinos from nearby radio galaxies could be done. These predictions could be tested, to some extent, with neutrino telescopes \citep{arguelles}.  However, UHECR from distant and more powerful AGN could not reach the Earth. At the same time, much higher level of activity in distant sources should result in much stronger neutrino and electromagnetic emission. In fact, the extragalactic $\gamma$-ray sky is dominated by distant, rather than nearby (i.e. within 50-100~Mpc distance) sources \citep{fermi_agn}. Most of those sources are blazars, which are a special type of radio galaxies with jets aligned along the line of sight.  It is possible that bright $\gamma$-ray emission from blazars originates from interactions of high-energy protons. 
 
 \gr s produced by high-energy proton interactions are absorbed by $\gamma\gamma$ pair production in interactions with soft photon fields. This $\gamma\gamma$ pair production results in additional injection of $e^+e^-$ pairs.  Electrons and positrons produced in this way, as well as those from the charged pion decays, loose energy via synchrotron and inverse Compton (IC) emission in the energy range that goes from radio to \gr s. The synchrotron-IC emission from the secondary \ee\ contributes to the observed broad-band emission spectrum of blazars and radio galaxies. In fact, high-energy proton interactions followed by electromagnetic cascades might, in principle,  supply all the high-energy electrons responsible for the observed broad-band electromagnetic emissions from these sources. 

Even if the multi-wavelength measurement of the spectral energy distribution (SED) of sources is becoming more and more precise, it is difficult to distinguish in the high energy region
the primary accelerated \ee\ from the secondary \ee\ induced by proton-initiated cascades. 
A direct verification of the hadronic origin of blazar and radio galaxy activity could be achieved by the observation of high-energy neutrinos from charged pion decays. The energy output of pion production reactions is approximately equally divided between neutrino and electromagnetic/leptonic channels (see e.g. \citet{kelner_pp,kelner}). 
The electromagnetic+leptonic part develops into an electromagnetic cascade that transfers all source energy to radio-to-\gr\ radiation. On the contrary, neutrinos freely escape from the source without further interactions. The total power of neutrino emission is, therefore, comparable to the power of radio-to-\gr\ emission assuming a common hadronic origin. At the same time, the relation between neutrinos and parent protons is easily calculable, and the neutrino component is expected to have energies systematically higher than the electromagnetic component.  

If the blazars and radio galaxies emitted power comes from high-energy proton interactions, the neutrino flux could be detected by very-high-energy (VHE) neutrino detectors, like the recently completed IceCube telescope at the South Pole. The published upper limits on the neutrino flux from point sources in the Northern hemisphere are derived using one year of about one half of the detector (IC-40) \citep{ic40_limits}. The IC-40 limits and sensitivities in the energy region $\sim 10$~TeV$ -10$~PeV are already comparable (in terms of the energy flux) with the electromagnetic fluxes of the brightest blazars observed by the Large Area Telescope (LAT) on board of Fermi satellite in the 0.1-100~GeV energy band \citep{fermi_agn}. 

Unfortunately, the electromagnetic measurements cannot provide information on the expected shape of the neutrino spectrum, as well as on the energy range in which the neutrino flux is expected to appear. Indeed, this energy range can be determined only if the shape of the spectrum of high-energy protons accelerated in the source is known. In the case of $p\gamma$ interactions, the shape of the neutrino spectrum is also determined by the energy threshold of the photo-pion reactions, which depends on the energy of the soft photons with which the high-energy protons interact. 

The IC-40 discovery potential would be at the level of the neutrino signal from the proton-powered blazars if the neutrino spectral energy distribution would have a maximum in the energy band 10~TeV -- 10~PeV, where the detector has maximum sensitivity. The non-detection of neutrino emission from the brightest blazars  constrains the parameter space of hadronic models, specifically the proton spectrum and/or of the seed photon field parameters.

In what follows we discuss the constraints on the hadronic models that could be imposed by the combination of simultaneous IC-40 neutrino and LAT \gr\ observations. We select for our analysis a set of brightest blazars in the sky and explore the parameter space of $pp$ and $p\gamma$ models (Sec.~\ref{sec:models}, Sec.~\ref{sec:sources} and
Sec.~\ref{sec:constraints}). We consider only the minimal possible set of model parameters and define the excluded parameter regions
(Sec.~\ref{sec:constraints}). In the Appendix we explain how we estimate the IC-40 sensitivity. 

\section{Hadronic models of blazar activity}
\label{sec:models}

The blazar SED is dominated by radiation from non-thermal particle distribution in the jet. It is composed by two broad components. The low energy component is generally attributed to synchrotron radiation from electrons and positrons in the knots moving at relativistic speed in the jet. The high energy component can have different origins depending on the model considered: inverse Compton emission via up-scattering of the synchrotron photons (synchrotron-self Compton model \citep{bloom96}), inverse Compton emission via up-scattering of the external radiation photons (external Compton model \citep{mandau}), emission from decays of neutral pions produced in proton-proton interactions~\citep{nellen93,bednarek97,beall99,sibiryakov08,neronov09,barkov10}, proton-photon interactions \citep{mannheim92,mannheim93,bednarek99,proton_synchrotron,atoyan01,atoyan03,neronov_semikoz} or proton synchrotron radiation~\citep{proton_synchrotron,aharonian00,mucke03}.  The maximal attainable energy of \gr\ emission is limited either by the maximal energies of electrons or by the onset of $\gamma\gamma$ pair creation~\citep{sikora87}. 

The high-energy electrons responsible for the synchrotron and inverse Compton emission could be either directly accelerated in the jet or have a non-acceleration origin as secondary particles produced by protons and \gr s interactions. The latter case is the so called ``purely hadronic'' model of blazar activity in which all the power of electromagnetic emission from the source is supposed to originate from the interactions of high-energy protons. 
Their energy could be converted into electromagnetic power via three different channels: $pp$ or $p\gamma$ interactions or direct synchrotron emission.  The three channels might, in fact, operate simultaneously in a competing way but in the following we will make the simplifying assumption that one of the channels dominates. 

\subsection{Proton-proton interactions}

The observed blazar spectra could be formed as a result of relativistic proton interactions with the matter in the source. These $pp$ interactions could efficiently remove energy from the accelerated protons if the matter density in the source is high enough.  In the process of inelastic collisions with low-energy background protons ($p+p \rightarrow \pi^{\{+,-,0\}}+X$), the cross section of the pion production is almost energy independent, $\sigma_{pp} \sim 4\cdot10^{-26}\mbox{cm}^{2}$~\citep{begelman90,kelner_pp}.  Secondary particles produced in $pp$ interactions are neutrinos, \ee and \gr s. The cross-section of this process has a threshold given by, $E_{th}=m_p+m_\pi(m_\pi+4m_p)/2m_p\simeq 1.2$~GeV. Therefore most of the accelerated protons could interact in the source and, because of this, the spectra of the produced secondary particles approximately follow the shape of the parent proton spectrum.

The efficiency of $pp$ interactions is determined by the ratio of the distance covered by the high-energy protons during the escape time form the source, $D_{esc}$, and the proton mean free path, $\lambda_{pp}\simeq  3\times 10^{15}\left[n_p/10^{10}\mbox{cm}^{-3}\right]^{-1}$~cm, where $n_p$ is the density of the background non-relativistic matter. The structure of the accretion flow in BL Lacs is, most probably, significantly different from that in the Flat Spectrum Radio Quasars (FSRQ). The parent population of BL Lacs, the Fanaroff-Riley type I radio galaxies, are known to accrete at low, largely sub-Eddington rates, so that the accretion flow is radiatively inefficient. In the case of the proton beam propagating at the speed of light through the radiatively-inefficient accretion flow (RIAF) in BL Lacs, the radial density profile of the accretion flow scales with distance as $n_p\sim D^{-1/2}$. Hence, the mean free path of high-energy protons grows as $D^{1/2}$ leading to the fact that most of the energy release of the proton beam takes place at large distances, since the efficiency of the $pp$ interactions scales as $D_{esc}/\lambda_{pp}\sim D^{1/2}$ \citep{sibiryakov08}. 
On the other hand, in the case of FSRQ, where the accretion proceeds at nearly the Eddington rate, since there is no model predicting the proton target distribution outside the accretion disk, the radial density profile of the accretion flow outside the accretion disk is not well constrained. Protons might interact if they cross the dense accretion disk \citep{nellen93} or face an obstacle in the form of a star or a dense blob of matter \citep{bednarek97,barkov10}.

If the spectrum of accelerated protons is softer than $dN_p/dE\sim E^{-2}$, most of the electromagnetic and neutrino power output from the $pp$ interactions is contained in the energy band around $\sim 1$~GeV. In particular, strong $\pi^0$ decay \gr\ emission contribution should be present in the observed \gr\ spectrum in Fermi energy band. The contribution from the electromagnetic cascade initiated by the $\gamma\gamma$ pair production in the source and from the inverse Compton emission from the secondary $e^+e^-$ pairs from the $\pi^\pm$ decays could be at most comparable to the $\pi^0$ decay component.  We call this case ``GeV \gr\ flux dominated by $\pi^0$-decay". In this case  the \gr\ data provide information on both the normalization and the slope of the secondary neutrino spectrum. 

To the contrary, if the high-energy proton spectrum has the slope $dN_p/dE\sim E^{-\gamma_p}$ with $\gamma_p<2$, most of the energy output from proton interactions is at the highest energies, much above $\sim 1$~GeV energy band. In this case the \gr\ emission in the GeV band is not necessarily dominated by $\pi^0$ decay \gr s. Instead, the main contribution could come from the $e^+e^-$ pairs produced either in the $\pi^\pm$ decays of in the $\gamma\gamma$ induced cascade.   We call this case ``GeV \gr\ flux dominated by electromagnetic cascade" in the following. The overall power of the cascade emission provides a measure of the overall power of neutrino emission. The spectral properties of the \gr\ emission are not related to those of neutrino emission, so that no estimate of the parameters of the neutrino spectrum are available from the \gr\ data.

\subsection{Proton-$\gamma$ interactions}

The accelerated relativistic protons could also interact with the radiation fields. The target photon field could be the jet synchrotron emission or it can be external to the jet, e.g. originate from the accretion disk or from the jet/accretion disk radiation reprocessed in the Broad Line Region (BLR) or from the radiation of the dusty torus~\citep{reimer11}. The efficiency of the $p\gamma$ process depends on the density and distribution of the photon target \citep{begelman90}. The main energy loss of protons is via the neutral and pion production $N+\gamma\rightarrow N+\pi^{0}$, $p+\gamma\rightarrow n+\pi^{+}$, $n+\gamma\rightarrow p+\pi^{-}$.  In the laboratory frame, the proton threshold energy can be expressed as $E_{th}=(2m_pm_\pi+m_\pi^2)/(2\epsilon_{ph}(1-\beta cos(\theta)))$,  where $\theta$ is the angle between the direction of the photon with respect to the proton direction and $\beta$ the proton velocity.  In the case of an head-on collision and $\beta\rightarrow 1$ limit , this leads to the minimal required proton energy\footnote{We assume the unit system with $c = 1$.}:
\begin{equation}
\label{eq:threshold}
E_{th}=\frac{2m_pm_\pi+m_\pi^2}{4\epsilon_{ph}}\simeq 7\times 10^{16}\left[\frac{\epsilon_{ph}}{1\mbox{ eV}}\right]^{-1},
\end{equation} 
where $m_p$, $m_\pi$ are the masses of proton and pions respectively and $\epsilon_{ph}$ is the energy of the target photons. This process has a higher energy threshold than the $pp$ interaction.
Because of this energy threshold, only the highest energy protons could efficiently interact with the soft-photon fields. Neutrino and electromagnetic emission are consequently produced via pion decays, $\pi^0\rightarrow 2\gamma$, $\pi^\pm\rightarrow \mu^\pm+\nu_\mu(\overline \nu_\mu), \mu^\pm\rightarrow e^\pm+\nu_e(\overline\nu_e)+\overline\nu_\mu(\nu_{\mu})$.

In the $p\gamma$ models, the characteristic feature of the neutrino spectrum is that, independently of the primary proton spectrum shape, the neutrino spectrum extends from the threshold energy (Eq.~\ref{eq:threshold}) up to the maximal energies of the protons, $E_{max, p}$. The condition for non-null efficiency in the $p\gamma$ models is therefore that $E_{max, p} > E_{th}$. 

Contrary to the neutrino emission, \ee and \gr s initially produced in the energy range $E_{th}\lesssim E <E_{max,p}$, would not be able to escape from the source. Hence, they induce an electromagnetic cascade with degrading energy of the electromagnetic component down to the energy range at which the source becomes transparent with respect to the $\gamma\gamma$ pair production ($E < 1$~GeV$-1$~TeV). 

The efficiency of the $p\gamma$ interactions in the source is also determined by the ratio of the mean free path of protons through the soft photon background, $\lambda_{p\gamma}$, and the time of escape of protons from the source, $t_{esc}$. The mean free path can be expressed as $\lambda_{p\gamma} =\left(\sigma_{p\gamma} n_{ph}\right)^{-1}\simeq 10^{13}\left[\sigma_{p\gamma}/10^{-28}\mbox{cm}^2\right]\left[n_{ph}/10^{15}\mbox{cm}^{-3}\right]$~cm, where $\sigma_{p\gamma}$ is the $p\gamma$ cross-section and $n_{ph}$ the photon density. In sources accreting at nearly the Eddington rate (such as FSRQ) the dense soft photon field is produced by the accretion disk and the density can be estimated to be $n_{ph}=L_s/(4\pi R_s^2 \epsilon_{ph})\simeq 10^{14}\left[L_s/10^{45}\mbox{ erg/s}\right]\left[R_s/10^{15}\mbox{cm}\right]^{-2}\left[\epsilon_{ph}/10\mbox{ eV}\right]^{-1}$~cm$^{-3}$, where $L_s$ is the source luminosity and $R_s$ is the source size. Taking into account that the $p\gamma$ cross section takes values of approximately $\simeq 6\times 10^{-28}$~cm$^{2}$ at the $\Delta$ resonance and $\simeq 1\times 10^{-28}$~cm$^{2}$ elsewhere, one finds that even if protons escape from the source region at the light crossing time scale, $t_{esc}\simeq R_s/c$, $p\gamma$ interactions could efficiently remove power from the high-energy protons. 

In the case of the RIAF, a decomposition of the electromagnetic emission as function of the distance to the black hole can be made \citep{sibiryakov08}.
The infrared (IR) synchrotron radiation is produced close to the black hole, the inverse Compton interactions occur at a larger radii, producing IR/optical emission, and finally, the Bremsstrahlung X-ray emission, dominates still further away from the black hole, at distances up to $10^3$ Schwarzschild radii. Typically, the densest radiation field is the synchrotron IR emission concentrated toward the black hole horizon. This densest photon field could, in principle, have a density of $n_{ph}\simeq 10^{13}\left[L_s/10^{41}\mbox{ erg/s}\right]\left[R_s/10^{15}\mbox{cm}\right]^{-2}\left[\epsilon_{ph}/10^{-2}\mbox{ eV}\right]^{-1}$~cm$^{-3}$ which is sufficiently high for efficient $p\gamma$ interactions. However, the low energy of background synchrotron photons $\epsilon_{ph}\sim 10^{-2}$~eV requires very high-energetic protons around $\sim 10^{19}$~eV in order to produce efficient pion production (see Eq. \ref{eq:threshold}). 
Under these conditions, the bulk of the neutrino emission from $p\gamma$ interactions in BL Lacs is expected to be concentrated around energies of $\gtrsim 10^{19}$~eV, which are beyond the scope of the IceCube limit we are considering.

The distribution of the photon target will also influence the shape of the produced neutrino spectra. For an optically thick medium, for example, if
the high-energy protons interact with the broad-band continuum synchrotron radiation in the blazar jet~\citep{mannheim92},  the effective threshold for efficient $p\gamma$ interactions is determined not by the characteristic energy of the background photon field (as in the case of interactions with black-body-like emission from the accretion disk), but by the energy at which the 
 soft-photon field density becomes diluted enough to allow high-energy protons to escape from the source. 
 Alternatively, assuming a power law shape for the photons field with a steep photon spectrum, the process of pion creation can be dominated by the interactions close to the energy threshold (at the $\Delta$-resonance)  \citep{mucke99}. At energies above the characteristic energy, the neutrino spectrum is expected to be similar to the proton spectrum, due to the nearly 100\% conversion efficiency from proton power to neutrino and electromagnetic power. At energies below the characteristic energy, the shape of the spectrum is determined by the relation between the energy-dependent mean free path of protons and the distance covered by protons over the energy-dependent escape time. Different combinations of mean free path and of the escape time energy dependent effects might give rise to rather different (power law-type) neutrino spectra below the characteristic energy.

  In general, both the distribution of the high-energy protons and  the target photon field seen by the high-energy protons in the blazar central engine or jet are highly anisotropic. These anisotropies strongly affect the efficiency of $p\gamma$ interactions in the source and the energy threshold for the pion production. For example, if the high-energy proton beam is generated in the AGN central engine and propagates through the quasi-thermal radiation field produced by the accretion disk, the energy threshold for the pion production would scale with the distance $d$ from the central engine as $(d/R_{disk})^2$, where $R_{disk}$ is the size of the quasi-thermal emission region from the disk. The increase of the threshold energy is due to the fact that the collision angle between the high-energy protons and the disk photons would scale as $\theta\sim d/R_{disk}$. In the first approximation, the increase of the threshold could be equivalently described by the decrease of the characteristic energy/temperature  of the soft photons, i.e. by the substitution $\epsilon_{ph}\rightarrow \epsilon_{ph}(R_{disk}/d)^2$ in Eq. (\ref{eq:threshold}).  Otherwise, if the disk photons are efficiently scattered in the BLR, the proton beam would see an isotropic photon field (in the reference frame of the AGN central engine) and the threshold energy would not change.
 
  In a similar way, if high-energy protons interact with the synchrotron photons generated by the blazar jet moving with the bulk Lorentz factor $\Gamma$ in the same direction as the proton beam), the collision angle between the high-energy protons and jet photons is $\theta\sim \Gamma^{-1}$, so that the threshold for the pion production scales as $\Gamma^2$, compared to the threshold which would be found in the AGN (or observer) reference frame in the case of isotropic distributions of high-energy protons and soft photons. Similarly to the case of the disk radiation, the change of the threshold energy could be described by the substitution $\epsilon_{ph}\rightarrow \epsilon_{ph}/\Gamma^2$ in Eq. (\ref{eq:threshold}). The increase of the energy threshold for the pion production could be equivalently described in the reference frame comoving with the jet. In this frame the energies of the soft synchrotron photons and of the protons are by a factor $\Gamma^{-1}$ lower than in the AGN / observer reference frame. Applying the Eq. \ref{eq:threshold} to the jet frame proton and soft photon energies, one would find that the threshold energy in the laboratory frame scales as $\Gamma^2$.

     The following discussion of the constraints on the $p\gamma$ models is presented in the AGN central engine / AGN host galaxy frame, which is equivalent to the "observer" reference frame in the case of the low-redshift sources. 
To keep our results as independent of the (highly uncertain) anisotropy patterns of the high-energy proton and soft photon distributions, we  make a scan over different characteristic energies $\epsilon_{ph}$ (or temperatures $T\sim \epsilon_{ph}$) of soft photon radiation fields. This is equivalent to taking into account the different possible values of the factor $\Gamma$ and $(d/R_{disk})^2$ in the substitutions $\epsilon_{ph}\rightarrow \epsilon_{ph}/\Gamma^2, \ \epsilon_{ph}\rightarrow \epsilon_{ph}/(d/R_{disk})^2$ in Eq. (\ref{eq:threshold}).

 The parametrization of $p\gamma$ interaction cross-sections by \citet{kelner} applies for the isotropic target photon distribution, so that it is, strictly speaking,  not appropriate for the description of interactions of high-energy protons with  photons produced by the accretion disk or jet. However, the precise anisotropy pattern of the photon field in the possible regions of high-energy proton interactions in blazars is not known. One might attempt to derive it assuming particular models of the accretion flow, for example, the Radiatively Inefficient Accretion Flow (RIAF) in BL Lac type sources \citep{sibiryakov08}, or the radiation from the accretion disk scattered in the Broad Line region in the case of the Flat Spectrum Radio Quasars \citep{atoyan01} etc.  However, the uncertainties of such calculation and numerous model parameters / assumptions introduced in this case would not make the results more precise. Calculations done assuming isotropic target photon field could be used for the purpose of the order-of-magnitude estimate of the overall power of neutrino emission.

As far as the target photon spectra are concerned, we concentrate first on the thermal spectra characterized by a single parameter, the  temperature $T$. To test the dependence of our constraints on the hadronic model parameters on the shape of the soft photon spectrum we perform a more detailed analysis for the source 3C273, which provides the tightest constraints. For this source we consider also the characteristic accretion disk photons spectrum, given by the "big blue bump" in the observed UV spectrum of the source. Alternatively, we consider the soft photon field generated by the synchrotron emission in the jet, taking also the observed synchrotron (radio-to-X-ray) spectrum of the source (Fig.~\ref{fig:3C273_SED}).

\subsection{Proton/muon/pion synchrotron models}

Proton acceleration to high energies requires strong electromagnetic fields. At the same time, the presence of strong magnetic fields leads to inevitable synchrotron energy losses, which could efficiently remove energy from the high-energy protons and result in \gr\ emission. This opens the possibility that the observed \gr\  emission from the blazars might be the synchrotron radiation of the accelerated  protons. 

For instance, in a study made by \citet{aharonian00} for the case of Mrk501, the observed \gr\ emission could be explained by the synchrotron radiation of protons if the compact region of size $R \sim 10^{15}$~cm is magnetized with magnetic fields of the order of $B\sim30-100$~G, and if the energy of the protons exceeds $E_p\gtrsim 10^{19}$~eV.  In this case, the synchrotron energy loss time for protons, $t_{s,p}\sim 5\times 10^4\left[B/100\mbox{ G}\right]^{-2}\left[E_p/10^{19}\mbox{eV}\right]^{-1}$\mbox{ s}, could be comparable or shorter than the energy loss time via competing $pp$ and $p\gamma$ interactions mechanisms. The characteristic energy of proton synchrotron photons would be $\epsilon_{p,s}\sim 1\left[B/100\mbox{ G}\right]\left[E_p/10^{19}\mbox{ eV}\right]^2\mbox{ GeV}$. 

If the magnetic field strength in the source is larger than several tens of Gauss, the synchrotron emission dominate over the other interaction channels. Then,  since muons, and, possibly, pions would loose energy via synchrotron radiation before decaying, the efficiency of neutrino production will be reduced with respect to the case of $pp$ and $p\gamma$ models. Consequently, contrary to the $pp$ and $p\gamma$ scenarios, the flux of neutrinos emitted by the source is in this case not directly related to its \gr\ flux.

\section{Selection of blazars using Fermi data }
\label{sec:sources}

For this study we analyzed simultaneous observations of the brightest blazars in the 0.1-100~GeV \gr\ band by the Large Area Telescope (LAT) on board of Fermi satellite \citep{atwood09} in order to derive the expected level of neutrino flux. 
Regular data taking with LAT started on August 4, 2008. 
We consider also LAT data taken during the period August 4, 2008 -- May 9, 2009, i.e. simultaneous with the IC-40 observations~\citep{ic40_limits}. 

A preselection of blazars from the two-year Fermi catalog~\citep{fermi_catalog} was done by requiring an average flux (over the 0.1 - 100~GeV) greater than $10^{-10}$~erg/cm$^2$s.  For each selected blazar, we performed the standard Fermi data analysis using the Fermi Science Tools\footnote{http://fermi.gsfc.nasa.gov/ssc/data/analysis/scitools/} version v9r23p1. We filtered the data using the {\tt gtselect} tool to select only events which are most likely \gr s ({\tt evclass=3}). All the sources considered in our analysis are at high Galactic latitudes $|b|>10^\circ$ and therefore the diffuse sky background around the source positions is not strongly variable on the degree scales. Taking this into account, we used the "aperture photometry" method to calculate the lightcurves and spectra of the sources. We have verified that the aperture photometry method of spectral extraction agrees with the results of the likelihood analysis (an alternative spectral extraction method used for the LAT data analysis).  The exposure is calculated using the {\tt gtexposure} tool also part of the Fermi framework. 

\subsection{Lightcurves of selected blazars}

\begin{table}
\small
\begin{tabular}{llccc}
\hline
Name & Type & $z$ & $F_{\rm IC40-period}$ & $F_{\rm 2-years}$\\
\hline
3C 66A           &   BL Lac &0.444$*$ & 17.7 & 25.8\\
AO 0235+164&  BL Lac  &0.94   &  24.3 & 16.5\\
Mrk 421           & BL Lac &0.030 & 30.1 & 30.8\\
3C 273            & FSRQ&0.158 & 21.2 & 11.2\\
PKS 1502+106& FSRQ&1.839 & 39.8& 30.9\\
B2 1520+31   &  FSRQ&1.487 & 16.1 & 13.3\\
PG 1553+113 &  BL Lac& -   & 19.8 & 17.2\\
BZB J2001+4352 &  BL Lac&  - & 11.3 & 10.0\\
3C 454.3           & FSRQ& 0.859 & 49.2 & 86.3\\
\hline
\end{tabular}
\caption{Selected sources for the northern hemisphere (declination $\delta>0^{\circ}$) with average fluxes $F_{\rm 2-years} > 10^{-10}$~erg/cm$^2$s in the 2-year Fermi catalogue ~\protect\citep{fermi_catalog}.
$F_{\rm IC40-period}$ is the average flux over the IC-40 period. Both fluxes are in $10^{-11}$ erg/cm$^2$s units.\newline
$*$ From SIMBAD database. The source redshift is uncertain, see \cite{3c66_redshift}.} 
\label{table:north}
\end{table}
%

\begin{figure*}
\begin{tabular} {c c c}
	\includegraphics[width=0.3\linewidth]{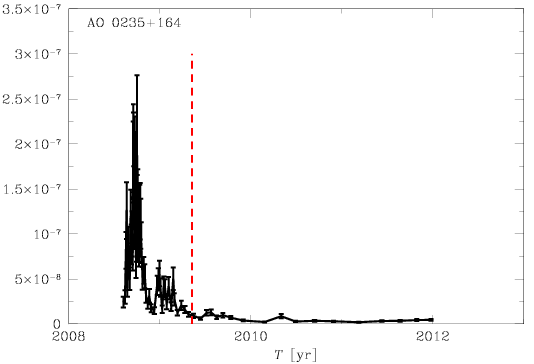}
	\includegraphics[width=0.3\linewidth]{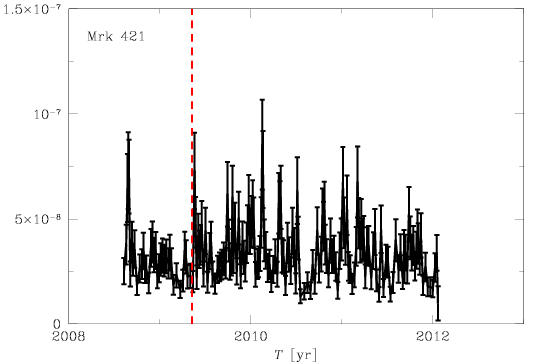}
	\includegraphics[width=0.3\linewidth]{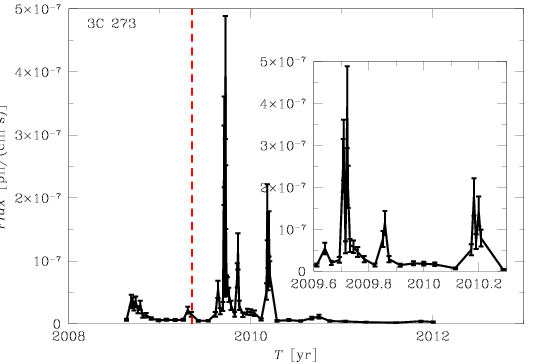}\\
	\includegraphics[width=0.3\linewidth]{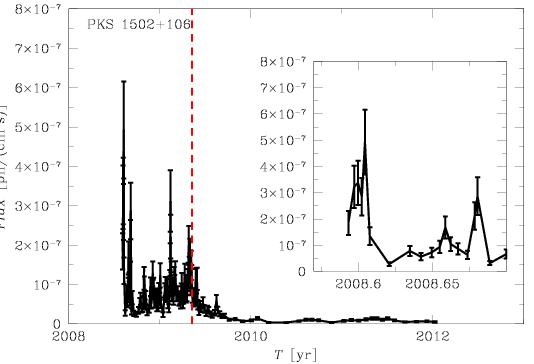}	
	\includegraphics[width=0.3\linewidth]{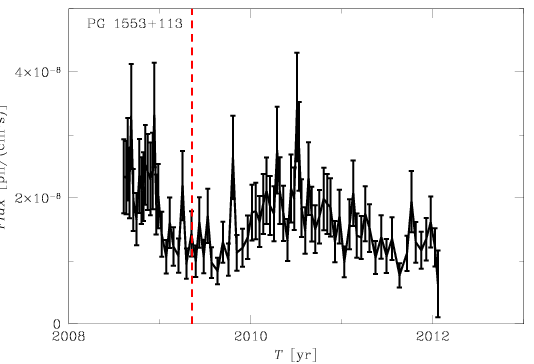}
	\includegraphics[width=0.3\linewidth]{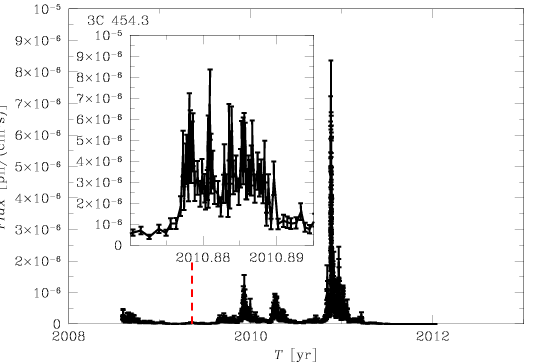}\\
\end{tabular}
\caption{LAT lightcurves of some of the selected blazars  (Tables \ref{table:north}) in the energy band above 1 GeV.  Vertical dashed line shows the end date of IC-40 exposure.}
\label{fig:lc}
\end{figure*}

Table~\ref{table:north} shows the list of northern sources with an average (over 2 years covered by the second Fermi catalog \citep{fermi_catalog}) flux greater than $> 10^{-10}$~erg/cm$^2$s. 
To verify that the flux of the selected sources was also high during the period of IC-40 observations the light curves of each source was produced since blazars are known to have highly variable \gr\ emission. Figure~\ref{fig:lc} shows the light curves for the selected sources.

Some sources, like AO 0235+164 and PKS 1502+10, were particularly active during the period of interest (indicated by the vertical red dashed lines in  Fig. \ref{fig:lc}). However, the presence of bright flares does not affect significantly the average flux during the IC-40 observation period, which is higher than the average flux by just a factor $1.1-1.5$  (see Tab.s \ref{table:north}). This means that the initial selection of the bright blazars based on the high two-year average fluxes already provides a representative `brightest blazar" set for our period of interest.

The sources listed  in Tab. \ref{table:north} belong to two different sub-classes of the blazar population: BL Lacs and FSRQ. As explained above, the properties of accretion flows in these two types of objects are different, therefore the efficiency of $pp$ and $p\gamma$ interactions in the two types of objects could also be very different. It is not clear in advance in which of the two types of objects the hadronic interactions are more efficient. BL Lacs and FSRQs are characterized by different spectra in the \gr\ range. 

\begin{figure}
\includegraphics[width=\columnwidth]{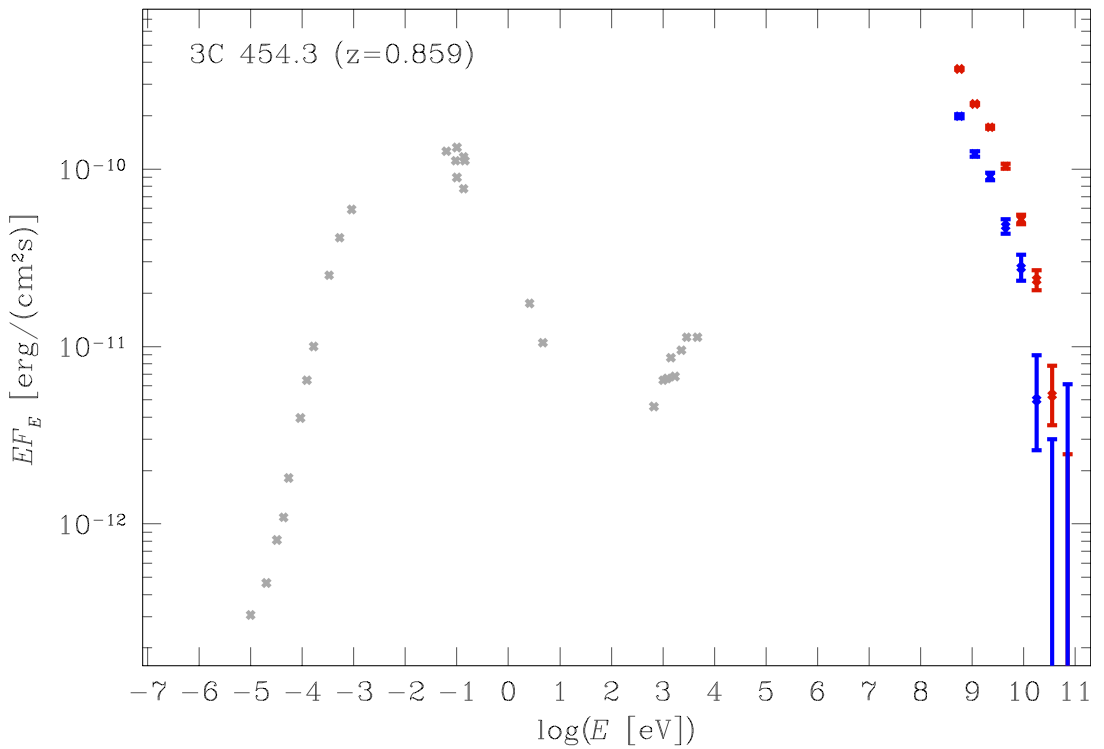}
\caption{Fermi data in the context of the broad-band spectral energy distribution of 3C 454.3. Blue color shows the spectrum during the period of IC-40 observations, red is the average over three-year Fermi data spectrum. Grey data points are from \citet{3c454_fermi}.}
\label{fig:3C454_SED}
\end{figure}

The Fermi measurement of the source flux in the 0.1-100~GeV range is an underestimate of the overall electromagnetic luminosity since it covers only 3 decades of the overall electromagnetic spectrum ($N_{E,Fermi}\simeq 3$). However, the underestimate is just by a factor of $C\sim N_{E,SED}/N_{E,Fermi}$ where $N_{E,SED}$ is the number of energy decades in which the energy flux is comparable to the flux in the Fermi band. Precise calculation of the factor $C$ is not possible because of the absence of coverage of the full electromagnetic spectrum. In the example of 3C 454.3, shown in Fig.~\ref{fig:3C454_SED}, one could roughly estimate $C \sim 2 - 3$. 

\begin{figure}
\includegraphics[width=\columnwidth]{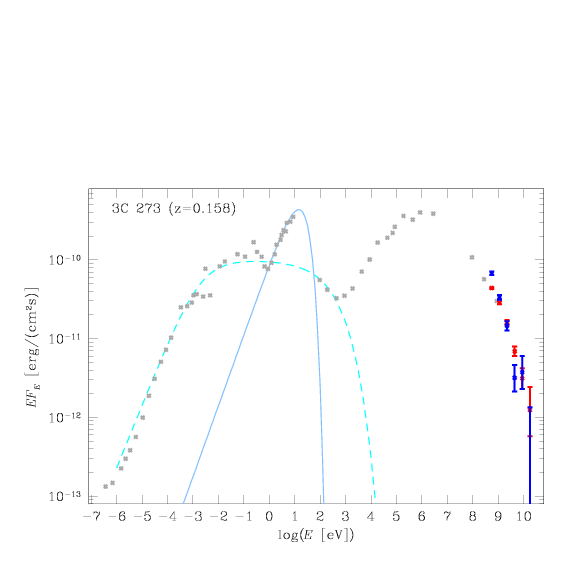}
\caption{Fermi data in the context of the broad-band spectral energy distribution of 3C 273. Red/blue data points are the same as in Fig.~\ref{fig:3C454_SED}. Grey data points are from \citet{courvoisier}. The dashed curve represents the fit for the synchrotron emission, while the continuous curve is the contribution due to the accretion disk.}
\label{fig:3C273_SED}
\end{figure}

The only important exception for which the estimate of the source power based on Fermi data can be largely underestimated is the blazar 3C 273, for which the spectral energy distribution is peaked in the MeV region (see Fig. \ref{fig:3C273_SED}). For this source, the flux in the MeV region could be higher by a factor $\sim 5$ than in the GeV band, so that considering Fermi measurement standalone is possibly an underestimate of an order of magnitude of the total source power. However, observations of the source in the MeV band date back to COMPTEL \citep{3C273_COMPTEL}, i.e. more than ten years ago. No simultaneous measurements with IC-40 are available.

\begin{figure}
\includegraphics[width=\columnwidth]{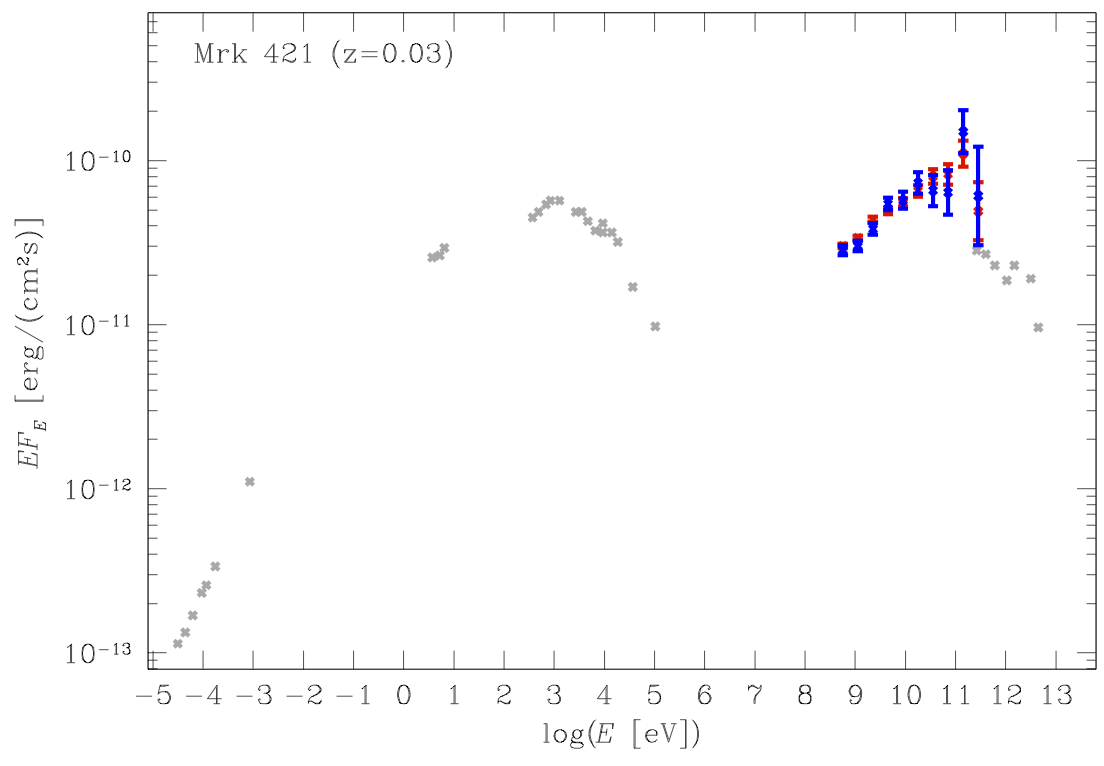}
\caption{Fermi data in the context of the broad-band spectral energy distribution of Mrk421. Red/blue data points are the same as in Fig.~\ref{fig:3C454_SED}. Grey data points are from \citet{fermi_mrk421}.}
\label{fig:Mrk421_SED}
\end{figure}

In the case of BL Lacs, the \gr\ dominance in the SED is less pronounced than in FSRQs (see Fig.~\ref{fig:Mrk421_SED} for an example of Mrk 421). In this case the power contained in the lower energy synchrotron component in the infrared-to-X-ray could provide a significant contribution to the overall source power. In the particular example of Mrk 421 this implies that $C\sim 2$ would be a good estimate for the factor by which the Fermi measurements under-estimate the total electromagnetic energy output of the source. 

\begin{figure}
\includegraphics[width=\columnwidth]{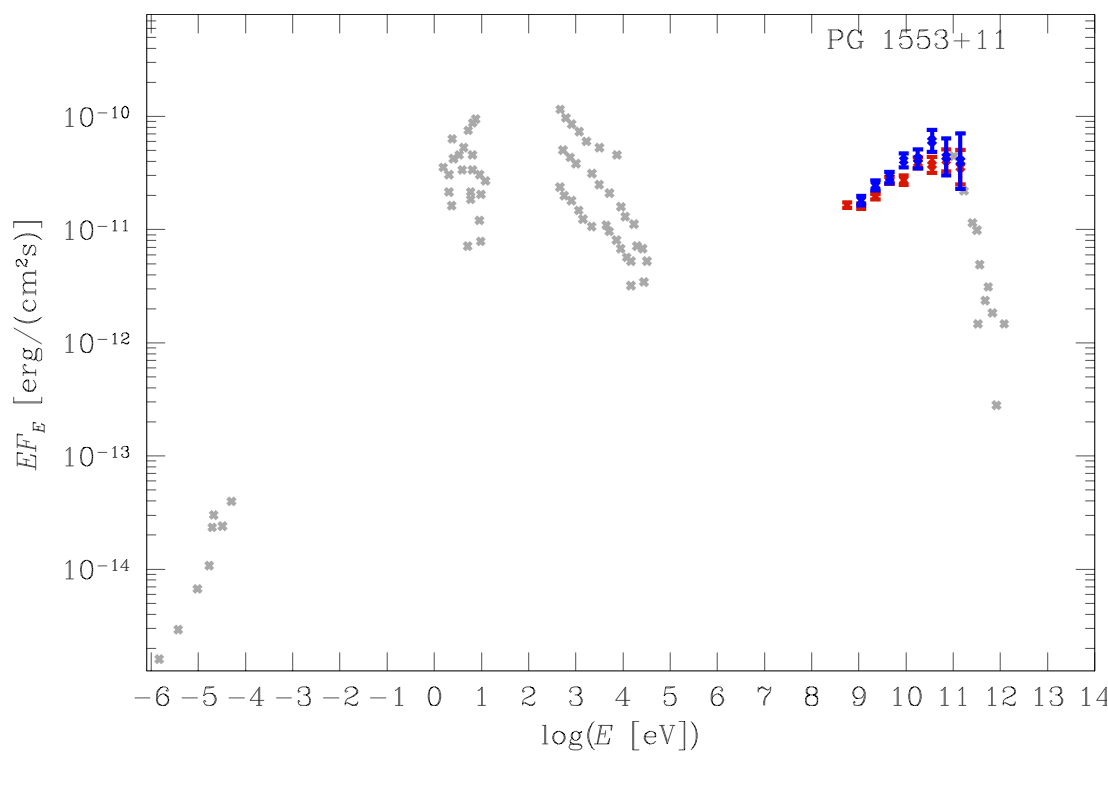}
\caption{Fermi data in the context of the broad-band spectral energy distribution of PG1553+113. Grey data points are from \citet{pg1553_fermi}. Red/blue data points are the same as in Fig.~\ref{fig:3C454_SED}.}
\label{fig:PG1553_SED}
\end{figure}

Any attempt to estimate the value of $C$ more precisely faces a difficulty related to the variability of the broad-band emission spectrum. A good example in this respect is PG 1553+113, for which the synchrotron emission component exhibits strong variability, with the overall power changing by at least a factor of three, see Fig.~\ref{fig:PG1553_SED}.

In the absence of precise measurements of the factor $C$, we adopt a conservative (under)estimate of the overall power of electromagnetic emission from the blazars listed in Table \ref{table:north}, based on the Fermi data only. We use the Fermi-measured flux to estimate the minimal possible neutrino emission power from the sources. The fact that the Fermi measurement always under-estimates the overall electromagnetic power by at least a factor of $C\sim 2$ implies that, in fact, the constraints on the purely hadronic models of activity of blazars, imposed by the IC-40 data are still stronger than those derived below.

 Another under-estimate of the total $\gamma$-ray power of the sources comes from the suppression of the flux in the energy band above 100 GeV due to the absorption on the Extragalactic Background Light (EBL) \citep{franceschini}. The effect is particularly important for the sources in which most of the power is concentrated toward the highest energy band $E>100$~GeV, like e.g. Mrk 421 and PG 1553+113. The 0.1-100 GeV band fluxes of Mrk 421, PG 1553+113 and other blazars listed in Table 1 are only slightly affected by the effect of absorption on the EBL.  However, ignoring the correction due to the EBL absorption in the estimate of the gamma-ray flux leads to a slight under-estimate of the neutrino power and therefore the constraints on the hadronic models of activity of blazars presented here are conservative.

\section{Constraints on hadronic interaction models}
\label{sec:constraints}

In the case of  the ``\gr\ dominated by electromagnetic cascade" models, which can occur for both the $pp$ and the $p\gamma$ interactions, gamma-ray data could be used to estimate the overall neutrino flux emitted from the sources; while in the case of  ``\gr\ dominated by $\pi^0$ decay"  models, which can only occur for $pp$ interactions, gamma-ray data can be used to estimate both the flux and the slope of the neutrino spectrum (see Section \ref{sec:models}). In this section we use these estimates to work out predictions for the neutrino signal in the IC-40 detector and compare these predictions with the IC-40 sensitivity. Non-detection of the sources in one-year exposure by IC-40 implies non-trivial bounds on the parameters of the high-energy proton spectra in the context of purely hadronic models of blazar activity. 

Note that such analysis would not be appropriate for testing the validity of purely hadronic models and their energetic (in)efficiency. It may well be that parameters of the source are such that interactions of high-energy protons in the source are inefficient (energy attenuation length is much longer than $ct_{esc}$ where $t_{esc}$ is the escape time of the protons from the source). In this case most of the high-energy protons escape in the form of ultra-high-energy cosmic rays, instead of powering the electromagnetic and neutrino emission. In what follows we do not aim at testing the validity of the purely hadronic model itself, we just constrain the model parameters assuming that electromagnetic and neutrino emission is powered by the high-energy proton interactions. 

We focus on the simplest form of the primary proton spectrum, a cut-off power law, $dN_p/dE\sim E^{-\gamma_p}\exp(-E/E_{max,p})$ characterized by just two parameters: the slope $\gamma_p$ and the cut-off energy $E_{max,p}$. The spectra of neutrinos produced in interactions of protons with cut-off power law spectra are characterized by different slopes $dN_\nu/dE\sim E^{-\gamma_\nu}$, which are determined by the slopes of the proton spectra and by the energy dependence of the production cross-section. At the same time, the published IC-40 upper limits and sensitivities on the neutrino flux \citep{ic40_limits} are applicable only to the case $\gamma_\nu=2$. For the blazars selected for the analysis, a re-calculation of the IC-40 sensitivity derived for  $\gamma_\nu=2$ is needed. We do this following the method proposed by \citet{neronov09}, taking into account the known energy response of the IceCube detector and the known level of the atmospheric neutrino flux. The details of the calculation are given in the Appendix. 

We follow Ref.~\citet{kelner_pp} to calculate secondary neutrino and photon spectra produced during $pp$ interactions. The parametrization described in this paper is optimized in the energy range $10^{11}$~eV to $10^{17}$~eV. In the frame of this work, we use the parametrization over a larger energy band up to $10^{19}$eV. While for the secondary neutrino and photon spectra produced during $p\gamma$ interactions, we use ~\citet{kelner}. 

Pion production reactions produce neutrinos of different flavors with the ratio: $F_{\nu_e}:F_{\nu_{\mu}}:F_{\nu_{\tau}}=$2:1:0 in  the charged pion decay followed by the muon decay. Assuming maximal mixing, the ratio of the fluxes observed at Earth changes to: $F_{\nu_e}:F_{\nu_{\mu}}:F_{\nu_{\tau}}=$1:1:1 \citep{oscillations}. Since the sensitivity of the IceCube detector used in this study is for muon neutrinos, the all-flavor neutrino fluxes derived here are scaled to the ratio of muon neutrino at Earth by a factor of 1/3.

\subsection{ Constraints on the $pp$  model: $\pi^0$ decay dominated \gr\ spectrum}

In this section, we assume that the electromagnetic emission is dominated by the $\pi^0$ decay in the GeV band.  Within this hypothesis, we determine some constraints of the proton spectra.

If the \gr\ spectrum is dominated by  the $\pi^0$ decay component, measurement of the slope of the \gr\ spectrum by Fermi provides a measurement of the slope of the spectrum of high-energy protons. The slope and normalization of the neutrino spectrum could be then calculated in a straightforward way using the parameterized production spectra of neutrinos in $pp$ interactions. Extrapolating these neutrino spectra to the IceCube energy band one could compare the model predictions based on the \gr\ measurements with the IC-40 neutrino sensitivities.

\begin{figure}
\includegraphics[height=\columnwidth,angle=270]{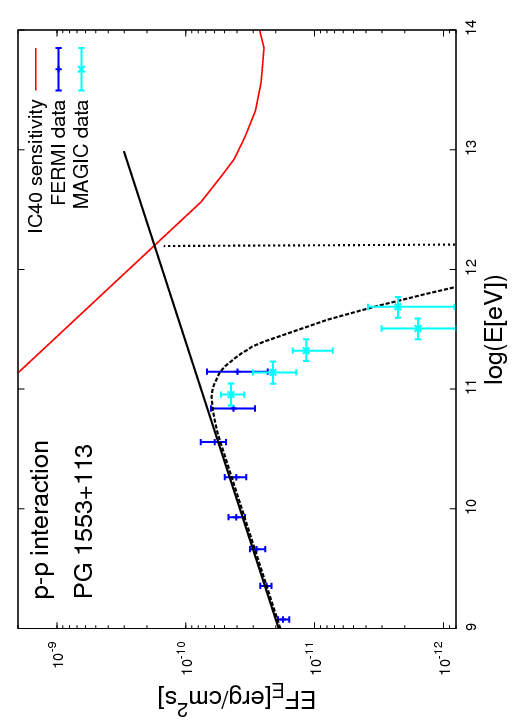}
\caption{\gr\ spectrum of PG 1553+113 recorded by Fermi (blue data points) during the IC-40 observations, compared with the IC-40 sensitivity. Cyan points are MAGIC data from \citet{pg1553_magic}. The red thick curve is the IC-40 sensitivity envelope. The vertical line shows the maximal energy to which the powerlaw type neutrino spectrum could extend.} 
\label{fig:pg1553}
\end{figure}

\begin{figure}
\includegraphics[height=\columnwidth,angle=270]{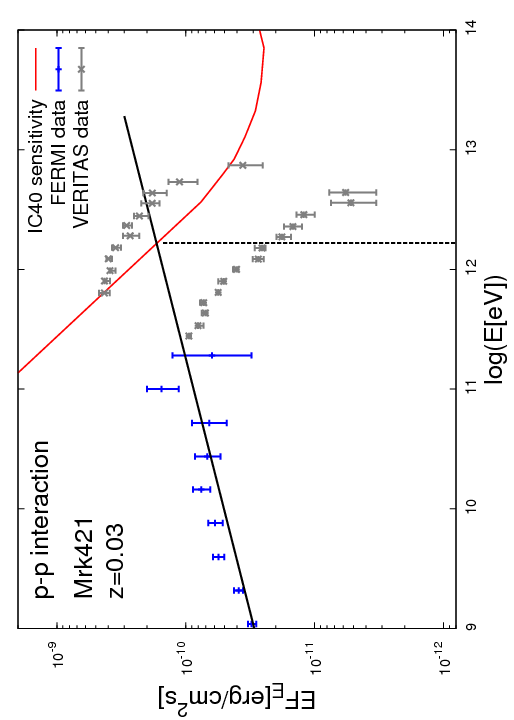}
\caption{Fermi observations (blue points) of Mrk421 during the IC40 period together with maximum and minimum fluxes observed by VERITAS (grey points) in 2008 \citep{veritas}. The vertical line shows the maximal energy to which the powerlaw type neutrino spectrum could extend. } 
\label{fig:Mkr421_pion0GeV}
\end{figure}

IC-40 impose restrictions on the cut-off energy of the proton spectrum in the case of sources with relatively hard \gr\ spectra. Two examples of such hard spectrum sources are shown in Figs. \ref{fig:pg1553} and \ref{fig:Mkr421_pion0GeV}. One could see that the spectra of these two BL Lacs are slightly harder than $\gamma_p=2$. Taking into account that the slope and normalization of neutrino spectra from $pp$ interactions approximately follow those of \gr\ spectra, one obtains the simplest estimate of the neutrino spectrum via a power law extrapolation of the Fermi spectrum into the domain of IC-40 sensitivity. Note that the suppression of the \gr\ flux at the highest energies is possibly due to the $\gamma\gamma$ pair production. In the case of PG 1553+113, shown in Fig. \ref{fig:pg1553}, the suppression is most likely due to the interactions with the EBL photons on the way from the source to the Earth (the dashed line shows the suppression for the assumed source redshift $z\simeq 0.5$) \citep{Danforth}.  Neutrino flux is not affected by the pair production and, therefore, could extend to higher energies without suppression. A power law extrapolation of the neutrino spectrum to the TeV-PeV energy band results in a neutrino flux higher than the IC-40 sensitivity, unless the neutrino spectrum has a high-energy cut-off. In both cases of PG 1553+113 and Mrk 421, the cut-off should be at $E_{max,p}\sim 1-10$ TeV.  In the case of Mrk 421, the IC-40 sensitivity is at the level of the measured \gr\ flux from the source in the TeV energy range. After a longer exposure, the full IceCube detector will have the potential to rule out the model in which the \gr\ emission is dominated by the pion decay component. Otherwise, if the TeV \gr\ emission is dominated by the neutral pion decay \gr s, 1-10~TeV neutrinos should be readily detected by the full IceCube.
 
Note that if the $\gamma$-ray emission from blazars is indeed dominated by the pion decays component, IC-40 results rule out the possibility that blazars (and, therefore, their parent population, radio galaxies) are accelerating protons up to the ultra-high-energy cosmic ray band ($\sim 10^{20}$~eV). This means that the hypothesis of UHECR origin in radio galaxies and blazars could be valid only if $\gamma$-ray emission from these sources is not produced directly in the decays of neutral pions.

\subsection{ Constraints on the $pp$  model: cascade dominated \gr\ spectrum}

Let us now derive constraints on the proton spectrum parameters assuming that the GeV emission is dominated by the electromagnetic cascade emission.

If the observed flux in the GeV-TeV range is dominated by the inverse Compton emission, rather than by the $\pi^0$ decay component, the spectral characteristics of the \gr\ signal could not be used in the estimates of the spectral characteristics of the neutrino signal. Only the overall normalization of the \gr\ flux provides the information on the level of the neutrino flux from the sources. The shape of the neutrino spectrum depends in this case on two model parameters: $E_{max,p}$ and $\gamma_p$ characterizing the primary proton spectrum.

\begin{figure}
\includegraphics[height=\columnwidth,angle=270]{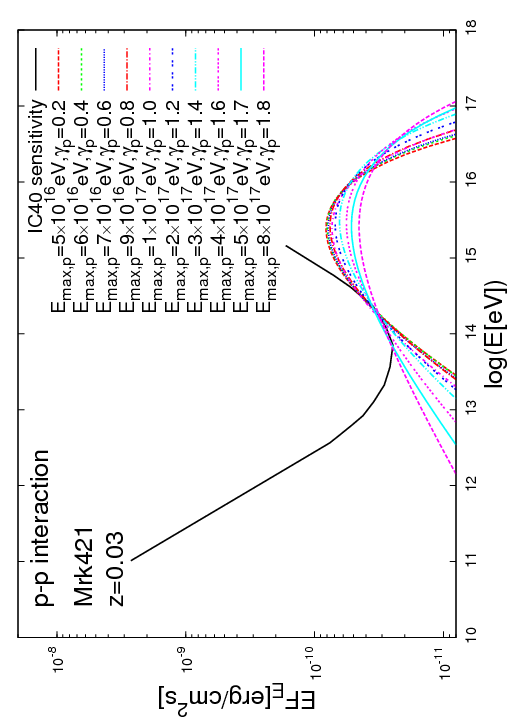}
\caption{Neutrino spectra expected for Mrk 421 for proton-proton interactions and different primary proton spectra characterized by the indicated proton cut-off energy, $E_{max,p}$, and by the 
proton spectral index, $\gamma_{p}$. The Fermi integrated flux is used for normalization. The black line is the envelope of power law sensitivities of IC-40. Only neutrino fluxes not exceeding the envelope sensitivity are shown. }
\label{fig:Mrk421.cascadeGeV}
\end{figure}

\begin{figure}
\includegraphics[height=\columnwidth,angle=270]{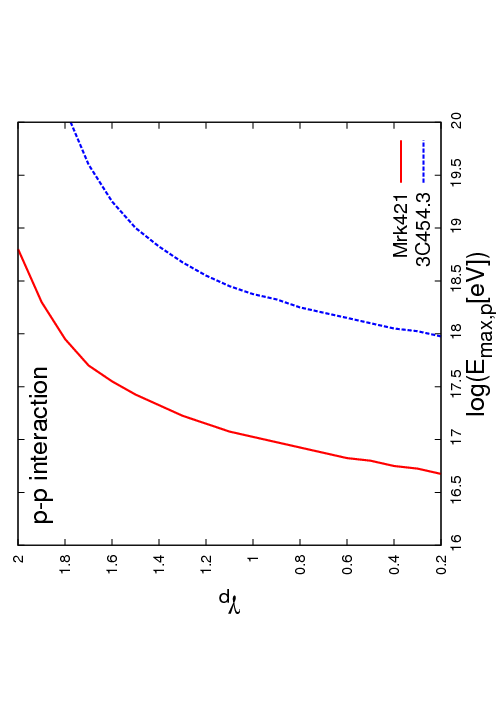}
\caption{Exclusion plot for $pp$ interaction model parameters (in the cascade dominated \gr\ emission case. The excluded range of parameters is to the left and above each curve. Constraints imposed by other sources are weaker. The constrains obtained for the sources Mrk 421 lie on the top and the one for 3C 454.3 lie on the bottom.}
\label{fig:pp_exclplot}
\end{figure}

A first  estimate of the allowed parameter space can be obtained by selecting, for each $\gamma_p$ value, the range of acceptable values of $E_{max,p}$ using the condition that the neutrino flux should not exceed the sensitivity limit of IC-40.  An example of the derivation of constraints on $\gamma_p, E_{max,p}$ is shown in  Fig~\ref{fig:Mrk421.cascadeGeV}. As the cross section for $pp$ interaction is almost constant, choosing larger values of $E_{max,p}$ would displace the bulk of neutrino production to higher energies while the choice of a softer spectral index ($\gamma_p$) will extend the neutrino emission to lower energy. 
Since the total emitted neutrino power is limited from below by Fermi electromagnetic observations, for a given  $\gamma_p$, the choosing a too low $E_{max,p}$ or too soft $\gamma_p$ would result in overproduction of neutrino flux in the energy range where IceCube is most sensitive so that the flux would exceed the IC-40 sensitivity. This implies that IC-40 data impose a lower bound on $E_{max,p}$ and an upper bound on $\gamma_p$.  Scanning cases for the two parameters $\gamma_p, E_{max,p}$ produces the exclusion plot in Fig~\ref{fig:pp_exclplot}, where two sources are considered: 3C 454.3 and Mrk421. The quasar 3C 454.3 is the brightest source of the sample (see Table \ref{table:north}). It results in the strongest constraints on both $E_{max,p}$ and $\gamma_p$. 

The lower bound of the allowed region ($E_{max,p} \gtrsim 10^{18}-10^{19}$~eV)
indicates that in the framework of $pp$ model with  cascade dominated \gr\ emission,  blazars should accelerate protons in the UHECR range: otherwise the model is ruled out by IceCube. Hence, a multi-year exposure of the full IceCube detector should be largely sufficient to either detect the neutrino signal from the blazars in the energy range above $\sim 10^{15}$~eV or completely rule out the $pp$ model of blazar activity. 

Note that the lower bound on $E_{max,p}$ does not depend on the efficiency of $pp$ interactions in the sources. Indeed, if the efficiency is low, most of the protons would escape from the source, transferring only a small fraction of their power to gamma-rays and neutrinos. However, the equal power re-distribution between the electromagnetic and neutrino components would hold also in this case.  In the context of purely hadronic models, the (in)efficiency of $pp$ interactions in the source could be estimated based on the observed variability properties of electromagnetic emission.  The power contained in high-energy protons could be efficiently converted into electromagnetic power through $pp$ interactions on a characteristic variability time scale $t_{var}$, if the mean energy loss time $t_{pp}=(K\sigma_{pp}n_p)^{-1}$ ($K$ is the inelasticity of $pp$ interactions, $n_p$ is the target proton density) is shorter than $t_{var}$. The requirement of efficiency of $pp$ interactions implies a lower bound on $n_p\gg(\sigma_{pp}Kt_{var})^{-1}$.  For example, in the case of Mrk 421, the minimal observed variability time scale is about 0.5 hr in the TeV energy band  \citep{gaidos96}. This imposes a constraint $n_p\gg 10^{12}\left[t_{var}/0.5\mbox{ hr}\right]^{-1}$~cm$^{-3}$.  This range of the densities is not unreasonable for the central parts of the RIAFs  \citep{narayan94,sibiryakov08}. In the case of 3C 454.3, the variability time scale in the Fermi energy band is $t_{var}\sim 1$~d \citep{ackermann10}. This implies a weaker lower bound on $n_p$, compared to Mrk 421. In 3C 454.3 the accretion flow is radiatively efficient and is not described by RIAF, but the typical density scales of the accretion flow in and outside the accretion disk in the blazar central engine are most probably much higher than in the case of RIAF.

\subsection{Constraints on the $p\gamma$ model}

Contrary to the $pp$ models which could be characterized by only two parameters, the ``minimal" $p\gamma$ model needs at least three parameters. The additional parameter describes the energy or temperature of the soft photons which serve as targets for the $p\gamma$ interactions. In the case of FSRQs, the soft photon field could be produced by the accretion disk, or by the BLR. As the spectrum of emission of BLR is formed by re-processing of the "big blue bump" spectrum, its spectrum is expected to be similar to that of the accretion disk. In those cases the typical photon energies are determined by the disk temperature $T$. On the contrary, in the case of photons coming from the jet, the soft photon spectrum has a power law shape, and hence the choice of the typical soft photon energy is more problematic. The appropriate definition of the characteristic soft photon energy is then the energy at which the soft photon field becomes dense enough to dump the high-energy proton power in the source. 

In the first approximation, we limit the analysis to the model where the soft photon field is provided by the accretion disk or the BLR and assume that the disk is characterized by a single temperature $T$,  this is a simplification of the realistic situation where the disk or BLR spectrum are a superposition of the multi-temperature black body spectra. A more detailed approach is followed for the case of 3C273. In this case we consider the interaction of the protons with more realistic target photon fields using the multiwavelength observations of this source  (see Fig.~\ref{fig:3C273_SED}).

\begin{figure}
\includegraphics[height=\columnwidth,angle=270]{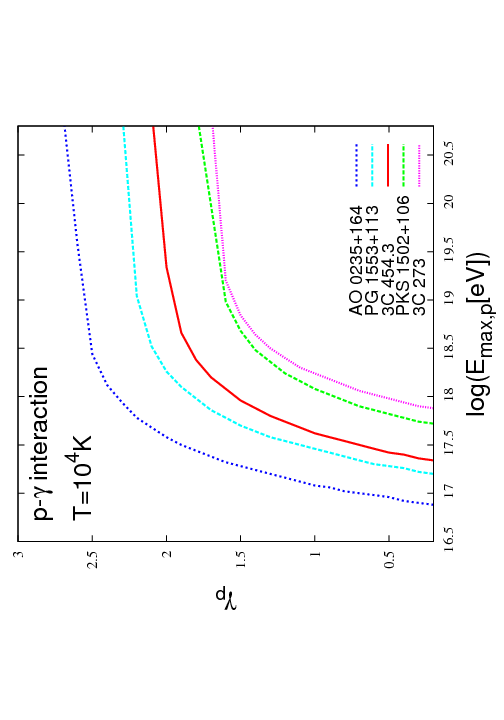}
\caption{Constraints on ($E_{max,p}, \gamma_p$) deduced in the $p\gamma$ model, assuming $T=10^{4}$K for different sources in Tab.~\ref{table:north}. The excluded regions lie above the curves. From top to bottom, the constrains are obtained for the sources: AO 0235+164, PG 1553+113, 3C 454.3, PKS 1502+106 and 3C 273. Best constraint is from 3C 273.}
\label{fig:pg_T1e4}
\end{figure}
%
\begin{figure}
\includegraphics[height=\columnwidth,angle=270]{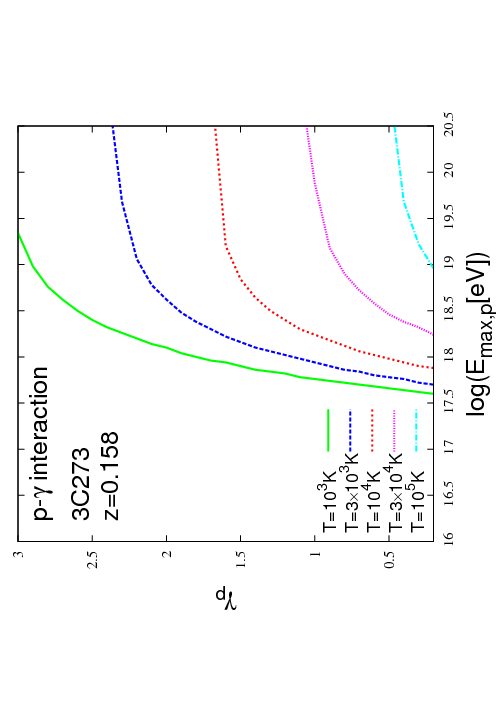}
\caption{As in the previous figure, but only for 3C 273, and for different temperatures of the soft photon field. From top to bottom, the lines represent the constrains obtained for the temperatures of: $T=10^{3}$K, $T=3\cdot 10^{3}$K, $T=10^{4}$K, $T=3\cdot10^{4}$K and $T=10^{5}$K. }
\label{fig:pg_exclplot}
\end{figure} 

\begin{figure}
\includegraphics[height=\columnwidth,angle=270]{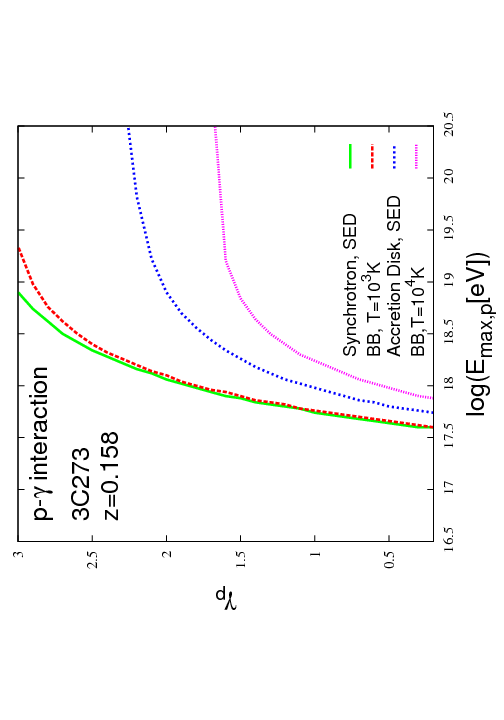}
\caption{Constraints on ($E_{max,p}, \gamma_p$) deduced in the $p\gamma$ model for proton interactions with different soft photon distributions. From top to bottom:  synchrotron photons with spectrum shown in Fig.~\ref{fig:3C273_SED},  black body (BB) of temperature $T=10^{3}K$, accretion disk spectrum from Fig.~\ref{fig:3C273_SED} and black body of  temperature $T=10^{4}K$. 
}
\label{fig:pg_exclplot_realistic}
\end{figure} 

The requirement that the level of the neutrino flux expected in the purely hadronic $p\gamma$ model should not exceed the envelope of the IC-40 power law flux sensitivities constrains the allowed range of the parameters $E_{max, p}, \gamma_p$ and $T$. Since the parameter space is three dimensional, the constraints could not be represented as a 2D exclusion plot, like in the case of the $pp$ model. Fig.~\ref{fig:pg_T1e4} shows cross-sections of this 3D allowed volume defined by the conditions $T=const$ for different sources in Tab.~\ref{table:north}. 

The strongest limit comes from the quasar 3C 273. 
Fig.~\ref{fig:pg_exclplot} shows the excluded region as function of the temperature for this source. For example, for a given temperature of the soft photon field $T=3\times 10^4$~K (corresponding to the typical temperatures of the accretion disks responsible for the ``big blue bump" UV emission, see Fig. \ref{fig:3C273_SED}),  the slope of the proton spectrum of 3C 273 is constrained to be harder than $\gamma_p\lesssim 1.7$. This slope of the emitting proton spectrum can be in the range expected for the shock acceleration models \citep{Schlickeiser}, if the distribution of protons injected by the acceleration process is not significantly modified by the effects of (energy dependent) escape from and attenuation in the source. Softer proton spectra are possible if lower temperature of the soft photon field is considered, see  Fig.~\ref{fig:pg_exclplot}. Lower soft photon temperatures also allow somewhat lower values of $E_{max,p}$. However, for all temperatures, $E_{max,p}$ remains in a range higher than $\sim 3\cdot10^{17}$~eV, implying that also in the model with $p\gamma$ interactions blazars have to be the UHECR sources for the model to be valid.

To test the robustness of the constraint on the cut-off energy of the proton spectrum, we investigated the dependence of the lower bound on $E_{max, p}$ on the shape of the soft photon spectrum for the particular case of 3C 273. The realistic spectrum of the accretion disk is a superposition of black body spectra with different temperatures, appearing as the "big blue bump" in the spectral energy distribution of the source, Fig. \ref{fig:3C273_SED}. The big blue bump spectrum in 3C 273 could be modeled as a cut-off powerlaw with the photon index close to $0$ and cut-off at $10$~eV. Repeating the analysis for such soft photon spectrum provides a constraint on the slope and cut-off energy of the proton spectrum shown in Fig. \ref{fig:pg_exclplot_realistic}.  As can be seen, the account of the realistic accretion disk radiation spectrum does not change the result qualitatively. In the particular case of 3C 273, the constraints on $E_{max, p}, \gamma_p$ found assuming the realistic accretion disk soft photon spectrum are equivalent to those found for the one-temperature black body spectrum with $T\sim 3\times 10^3$~K. Alternatively, we considered the soft photon spectrum generated by the synchrotron radiation in the jet, with the shape shown in Fig. \ref{fig:3C273_SED}. The constraint on $E_{max, p}, \gamma_p$ found under this assumption is also shown in Fig. \ref{fig:pg_exclplot_realistic}, which is equivalent to that derived assuming the blackbody soft photon spectrum with temperature $T\sim 10^3$~K. Therefore the details of the shape of the soft photon spectrum do not have a great impact on the rejection bounds that can be set on the basic parameters of the $p\gamma$ hadronic scenario.

The constraints on $E_{max, p}, \gamma_p$ depend on the anisotropy patterns of the high-energy proton beam and soft photon field. As it is mentioned in Section 2.2, the energy threshold of the photo-pion production depends on the geometry of the proton beam as well as on the location and geometry of the soft photon source. Dependence of the bound on $E_{max,\ p}$ on the bulk Lorentz factor of the jet $\Gamma$ or on the ratio of the typical size of the pion production region $d$ to the size of the accretion disk $R_{disk}$ could be found by the re-scaling of the characteristic energy/temperature  of the soft photons by $\Gamma^2$ and $(d/R_{disk})^2$, respectively.   

Similarly to the case of $pp$ the efficiency of $p\gamma$ interactions could be estimated based on  the variability time scale. In the particular case of 3C 273, one could find from Fig. \ref{fig:lc} that the variability time scale during the flaring activity of 3C 273 is about $t_{var}\sim 10$~d. The condition that proton energy loss time $t_{p\gamma}=(K_{p\gamma}\sigma_{p\gamma}n_{ph})^{-1}$ (where $K_{p\gamma}$ is the inelasticity, $\sigma_{p\gamma}$ is the cross section of $p\gamma$ interactions and $n_{ph}$ is the density of soft photons with energies above the pion production threshold) is shorter than $t_{var}$ imposes a constraint $n_{ph}\gg  3\times 10^{11}\left[t_{var}/10\mbox{ d}\right]^{-1}$~cm$^{-3}$. This soft photon density implies the luminosity due to the target photons of $L_{x}\gg 4\pi R^2n_{ph}\epsilon_{ph}\gg3\times 10^{43}\left[t_{var}/10\mbox{ d}\right] \left[\epsilon_{ph}/1\mbox{ eV}\right]$~erg/s, where $R=ct_{var}$ is the source size, $\epsilon_{ph}$ the typical soft photon energy and x the region from where the target photon originate. 

If the target photons originate from the accretion disk, we obtain that the required luminosity in the accretion disc should be of $L_{AD}\gg3\times 10^{43}\left[t_{var}/10\mbox{ d}\right] \left[\epsilon_{ph}/1\mbox{ eV}\right]$~erg/s and this estimate of luminosity scale is not unreasonable. The observed luminosity of 3C 273 is in the range of $\sim 10^{46}$~erg/s (see Fig. \ref{fig:3C273_SED}), so that in this particular case $p\gamma$ interactions could efficiently transfer the power from high-energy protons to photons and neutrinos, both during quiescent periods and during flares.

If the target photons originate from the jet, the above estimate should be understood in term of the quantities ($t_{var},\ L,\ R,\ n_{ph},\ \epsilon_{ph},\ ...$) defined in the jet comoving frame (denoted with $'$). To relate those quantities to those in the observer frame, one needs to take into account the Doppler factor (D) of the jet. In this case, with $t'_{var}=t_{var}D$, the minimum required photon target density in the comoving frame can be expressed as  $n'_{ph}\gg  3\times 10^{10}\left[t_{var}/10\mbox{ d}\right]^{-1}\left[D/10\right]^{-1}$~cm$^{-3}$. Assuming a jet Doppler factor of 10, the luminosity in the comoving frame corresponding to this photon density is $L'_{jet}\gg 4\pi R'^2n'_{ph}\epsilon'_{ph}\sim3\times 10^{43}\left[t_{var}/10\mbox{ d}\right] \left[\epsilon_{ph}/1\mbox{ eV}\right]$~erg/s, where  $\epsilon'_{ph}=\epsilon_{ph}\cdot D^{-1}$ is the photon energy and $R'=R\cdot D$ is the size of the emitting region, both expressed in the comoving frame. The observed luminosity corresponding to this minimal required luminosity is $L_{jet}=L'_{jet}D^{4}\gg 3\times 10^{47}\left[t_{var}/10\mbox{ d}\right]  \left[\epsilon_{ph}/1\mbox{ eV}\right][D/10]^4$~erg/s. This implies that the energy loss of the proton through $p\gamma$ interactions in the jet is efficient only for the brightest sources. 

\section{Discussion}

The constraints on the parameters of $pp$ and $p\gamma$ models of the activity of  blazars derived in the previous sections demonstrate the potential of the multi-messenger method to study astrophysical sources, which becomes available with the start of operation of IceCube neutrino telescope. Figs. \ref{fig:pp_exclplot} and \ref{fig:pg_T1e4}, \ref{fig:pg_exclplot} show that the combination of the one-year exposure of one half of the IceCube detector with the simultaneous \gr\ observations by the Fermi telescope can provide important constraints  on the possible high-energy proton acceleration in blazars in the context of hadronic models of blazar activity. 

In its simplest version, the hadronic model of blazar activity with $pp$ interactions is characterized by just two parameters describing the shape of the high-energy proton spectrum that can be constrained (see Fig. \ref{fig:pp_exclplot}). In this case IC-40 + Fermi data indicate that protons are accelerated at least to the energies of UHECR. Lower cut-off energies of the proton spectra would over-predict the fluxes of neutrinos in the TeV-PeV energy range where IC-40 is most sensitive. 

In the case of $p\gamma$ model, the limits imposed by the neutrino data depend on an additional parameter, which is not directly measured.  This additional parameter is the  typical energy of the soft photons. In this study we have limited our analysis to the simplest case when the soft photon field has a thermal spectrum, so that the characteristic energy of the soft photons is given by the temperature.  From Figs.~\ref{fig:pg_T1e4} and \ref{fig:pg_exclplot}, one could see that e.g. fixing the temperature of the soft photon field, one still finds relatively tight constraints on the slope and high-energy cut-off of the proton spectrum. For instance in the Figs. \ref{fig:pg_T1e4} and \ref{fig:pg_exclplot_realistic}, it is shown that when the temperature is fixed  at $10^4$K, the cut-off energies of the proton spectra are allowed  to be in the range  larger than $10^{17}-10^{18}$~eV for a sample of the brightest sources.  The slope of the proton spectrum is also constrained to be very hard. Changing the soft photon field temperature leads to modification of  these constraints, somewhat relaxing or strengthening the bounds on $E_{max,p}$ and $\gamma_p$. However, as it is shown in Fig. \ref{fig:pg_exclplot}, in the case of 3C 273, which provides the best constraints, the cut-off energy of the proton spectrum is always bound to be higher than $\sim 10^{18}$~eV.

The lower bound on the maximal proton energies derived from the neutrino data  provides an indirect constraint on the physical conditions at the acceleration site, in particular on the magnetic field strength $B$ and size $R$ of the acceleration site. This constraint stems from the  Hillas condition $E_{max,p}\ge e\cdot B\cdot R =10^{18}\left[B/10^4\mbox{ G}\right]\left[R/10^{12}\mbox{ cm}\right]$eV. The bound $E_{max,p}\gtrsim 10^{17}-10^{18}$~eV translates to the constraint $\left[B/10^4\mbox{ G}\right]\left[R/10^{12}\mbox{ cm}\right]>1$. This constraint is satisfied in several components of the blazar, in particular, in the central engine with typical size about the Schwarzschild radius of the black hole and strong magnetic field in excess of kG, in the parsec scale jet with $R\sim 1$~pc and magnetic fields possibly in the mG range and even possibly in the large kiloparsec scale jet with $R\sim 1$~kpc and $\mu$G magnetic fields.

It is interesting to note that the set of blazars providing the best constraints on the parameters of $pp$ and $p\gamma$ models is different. Tightest constraints on the $pp$ model parameters are given by the brightest Northern hemisphere blazars. This is not surprising, because the neutrino spectrum in the $pp$ model is expected to be emitted in a broad energy range, including the range in which the IceCube detector is most sensitive (from $\sim 10$~TeV to $\sim 10$~PeV). At the same time, the tightest constraints on the $p\gamma$ model parameters are imposed by the bright blazar(s) situated close to the horizon of IceCube detector (towards the Southern hemisphere). This is explained by the fact that in the $p\gamma$ models the neutrino flux is peaked toward the high-energies in excess of $\sim 10$~PeV. The sensitivity of IceCube at the highest energies improves around the horizon, so that the best sources for observations are bright blazars close to the equatorial plane.

The overall efficiency of proton interactions in the source in both $pp$ and $p\gamma$ models strongly depends on the  physical condition of the source, such as the size and location of the interaction region in the central engine, jet, BLR, etc. This in turn affects the neutrino flux which one expects from the source with a given proton power \citep{atoyan01,atoyan03}. However, because the (in)efficiency of proton interactions affects the gamma ray and the neutrino flux in the same way, our results are not sensitive to the efficiency of pp and p$\gamma$ interactions.

As it is mentioned in the Section 2.3, our study does not apply to the models in which the energy of the proton beam is used for the generation of proton or muon synchrotron emission. In this case the neutrino power of the source might be much smaller than the electromagnetic power. In fact neutrino power could be made arbitrarily  small in these scenarios, so that the constraints reported in this study are largely relaxed.

In the models with the cascade-dominated $\gamma$-ray emission, the observed electromagnetic power is generated by the secondary electrons only when their energy decreases below the pair production threshold. This implies that the electromagnetic emission might be delayed compared to the neutrino emission by the characteristic time of the cascade development. Another potential source of the time delay of electromagnetic emission is geometrical delay occurring due to the deflections of electrons and positrons by magnetic fields. If the electromagnetic cascade develops at the distance $d$ from the AGN central engine, in a cone with an opening angle $\theta\sim \Gamma^{-1}\sim 0.1$, the characteristic time delay is $d\theta^2/c\sim 1\left[d/30\mbox{ pc}\right]$~yr. Our study is based on the neutrino and $\gamma$-ray fluxes averaged on $\sim 1$~yr time scale. If the distance, $d$, is much larger than $\sim 30$~pc, the $\gamma$-ray and the characteristic time scale of activity of blazars is comparable to one year, and therefore the assumption that $\gamma$-ray flux is comparable to the neutrino flux might be violated. In this case the constraints on $E_{max, p}, \gamma_p$ derived above from the common $\gamma$-ray + neutrino data set might be relaxed.  

\section{Conclusions}

In this study, we investigated the constraints on parameters of purely hadronic models of activity of blazars imposed by the IceCube data. We assumed that the observed electromagnetic emission comes purely from high-energy proton interactions via $pp$ and/or $p\gamma$ channels and  used this assumption to estimate the level of neutrino flux which was subsequently compared to the IceCube sensitivity.  We neglected possible alternative energy loss channels which do not result in neutrino production, such as the proton synchrotron emission, and the synchrotron energy loss by muons and pions. We also did the simplifying assumption that the proton only interact via $pp$ or the $p\gamma$ interactions, where only one of the two mechanisms dominates. Taking into account the uncertainties of the anisotropy pattern of the soft photons fields in the blazar, we assumed that high-energy protons interact via $p\gamma$ interactions with approximately isotropically distributed soft photons. In our analysis, we have tried to systematically estimate the lowest possible expected neutrino power (e.g. restricting the estimate of electromagnetic luminosity to the gamma-ray band only, neglecting the effects of the absorption of the highest energy photons in interactions with EBL etc), to derive conservative constraints on the model parameters from the IceCube data.

An analysis of the hadronic models of blazar emission is performed in the previous sections and it is shown that IceCube is able to constrain these models. 
We find that one year long exposure with half of the detector can already set constraints on the possible shapes of the high-energy proton spectra and indicates that if the hadronic models are valid, blazars should be accelerating protons right up to the UHECR energies. 

This analysis indicates that harder spectra than $E^{-2}$ proton spectra and  cut off energies in the UHECR band  are favored by IceCube limits. In this ultra-high-energy band the atmospheric neutrino background is low and the detection of the sources would be possible even with single neutrinos coming from the direction of blazars. An obvious disadvantage of the signal concentrated at the highest energies is that the accumulation of a significant signal might require multi-year exposures. 

It is noticeable that the increase of the power output of high-energy protons leads to increased fluxes of both photons and neutrinos. Thus, the neutrino flux is expected to be higher during photon flares. Taking this into account, a dedicated analysis of extremely bright flares, like the flare of AO 0235+164 and other sources which happened during the IC-40 observation period, might increase the signal-to-noise ratio in the neutrino signal. Indeed, restricting the search of neutrino emission to the period of the bright flare the atmospheric muon and neutrino backgrounds are largely reduced (see Ref. \cite{mike}).
This should facilitate the detection of the signal from the source. We did not pursue such an approach in the present paper, where only the analysis of the time-average signal was done. We also notice that a large number of extremely bright flares from the sources listed in table \ref{table:north} has occurred after the end of operation of IC-40. During this time, more sensitive configurations of IceCube detector, with more strings, were in operation. The search for the neutrino emission accompanying bright \gr\ flares of blazars  would provide an improvement of the constraints on the model parameters, compared to those reported above. We leave such an analysis for future work.

\appendix
\section{Method to extract IceCube sensitivity curves}

The IceCube Neutrino Observatory is a neutrino telescope installed in the deep ice at the geographic South Pole. Its final configuration comprises
5,160 photomultipliers (PMTs) along 86 instrumented strings buried between 1.5-2.5 km in the ice. Its design is optimized for the detection of
high energy astrophysical neutrinos with energies above $\sim 100$ GeV.  The IceCube detector uses the Antarctic ice as detection volume where muon neutrino interactions
induced muons produce detectable Cherenkov light. The light propagates through the
transparent ice medium and can be collected by PMTs housed inside Digital Optical
Modules (DOMs). The DOMs are spherical, pressure resistant glass vessels each
containing a 25 cm-diameter Hamamatsu photomultipliers and their associated
electronics. 

Our constraints on the parameters of hadronic models are based on non-detection of neutrino flux from selected blazars in the IC-40 data \citep{ic40_limits}. The neutrino spectra expected in the hadronic models of blazars are hard in the energy range $1-10^3$~TeV where IceCube is most sensitive. At the same time, the IceCube publications, in particular \citet{ic40_limits}, conventionally report only upper limits on the much softer "standard" $dN_\nu/dE\sim E^{-2}$ type source spectra.  In this section we derive the upper limits on the neutrino flux from the sources with arbitrary powerlaw-type spectra, based on the knowledge of the detector characteristics. The detector characteristics relevant for the derivation of the upper limits for arbitrary powerlaw type spectra are the effective area $A_{eff}(\delta, E)$, which is a function of neutrino energy and of the source position on the sky, the cumulative point spread function (PSF) $f_{cum}(\psi,E)$, which is also a function of energy and the exposure time $T$. These characteristics are given by \citet{ic40_limits}. 

Knowing the spectrum of atmospheric neutrinos arriving from a declination $\delta$ in a solid angle $d\Omega$ \citep{honda2006},  $dN_{atm}(\delta,E)/(dEd\Omega)$, one could calculate the number of atmospheric neutrino events with energies above $E_{min}$, arriving in an angular bin of size $\psi$ during the observation time T:
\begin{equation}
\label{atmospheric}
N_{bgd}(E_{min},{\psi},T)=\pi\psi^2 T\int^{E_{max}}_{E_{min}} {A_{eff}(\delta,E)\frac{dN_{atm}(\delta,E)}{dEd\Omega}}dE.
\end{equation}
\citet{ic40_limits} give the values of $A_{eff}(\delta,E)$ at several reference declinations $\delta$. To calculate $A_{eff}$ at particular source positions, we interpolate $A_{eff}$ between two nearest reference declinations. 

Signal neutrinos arrive from a given direction on the sky within a circle of the radius $\psi$. The number of signal events within this circle is
 \begin{equation}
 N_{signal}(E_{min},\psi,T)=T\int^{E_{max}}_{E_{min}}A_{eff}(\delta,E)\frac{dN_{source}}{dE}f_{cum}(\psi,E)dE,
 \end{equation}
 where $dN_{source}/dE$ is the source neutrino spectrum. \citet{ic40_limits} give $f_{cum}(\psi,E)$ in two energy bins. We use either the high-energy or the low-energy PSF in the calculation of the upper limits for the soft spectra $\gamma_{\nu}>2$ and high-energy PSF for the harder spectra.  This is justified by the fact that the signal statistics in the two cases is dominated by the lower- and higher-energy events in the two cases, respectively.
 
The signal events are detected on top of the atmospheric neutrino background events (\ref{atmospheric}). To distinguish the signal from the atmospheric neutrino background, the signal level must be higher than the level of background fluctuations. The level of background fluctuations is  a function of the angle $\psi$ and minimal energy $E_{min}$. The data analysis should adjust the values of $\psi$ and $E_{min}$ in order to maximize the chances of detection of the signal on top of the background. Choosing too small $\psi$ would suppress the background, but also suppress the signal, so that the minimal detectable source flux would be high. Otherwise, choosing too large $\psi$ would result in the maximization of the source signal, but also in a too high level of the background, so that the minimal detectable source flux would also be high. An optimal choice of $\psi$ is achieved when most of the signal events are already contained in a circle of the radius $\psi$, but the number of background events within the same circle is still reasonably small. In a similar way, an optimal value of $E_{min}$ exists. The number of background neutrinos is rapidly decreasing with the increase of $E_{min}$, because the spectrum of atmospheric neutrinos is very soft. Thus, increasing $E_{min}$ suppresses the level of the background. At the same time, increase of $E_{min}$ also suppresses the source signal, possibly to a lower extent. An optimal choice of $E_{min}$ is that for which the signal is still high while the background is sufficiently suppressed. The optimal choice of $\psi, E_{min}$ depends on the slope of the source spectrum.
 
In order to find the optimal values of $\psi,E_{min}$ for each slope $\gamma_{\nu}$ of the source spectrum 
\begin{equation}
\label{source}
\frac{dN_{source}}{dE}=\Phi_0\left(\frac{E}{E_0}\right)^{-\gamma_{\nu}},
\end{equation}
($\Phi_0$ is normalization at an arbitrary energy $E_0$) we scan the parameter space $(\psi, E_{min})$ and compute the number of signal events $N_{signal}^{FC}(\psi,E_{min},T)$ which could be excluded at the 90\% confidence level,  given the number of background events in the angular bin of size $\psi$ above the energy $E_{min}$. We  use the method of \citet{FC} for this purpose. Having found $N_{signal}^{FC}(\psi,E_{min},T)$, we calculate the normalization $\Phi_0$ of the source spectrum (\ref{source}) corresponding to this event number, taking into account the exposure  $T$, the detector effective area $A_{eff}$ and the fraction of the source signal contained in a circle of the radius $\psi$.  The normalization for each spectral index of the particle distribution is
 \begin{equation}
 \label{n_s}
\Phi_0(\psi, E_{min})=\frac{{N_{signal}^{FC}(\psi,E_{min},T)}}{T\cdot f_{cum}(E,\psi)\cdot \int^{E_{max}}_{E_{min}}A_{eff}(\delta,E)\left(E/E_0\right)^{-\gamma_{\nu}}dE}.
\end{equation}

\begin{figure}
\includegraphics[height=\columnwidth,angle=270]{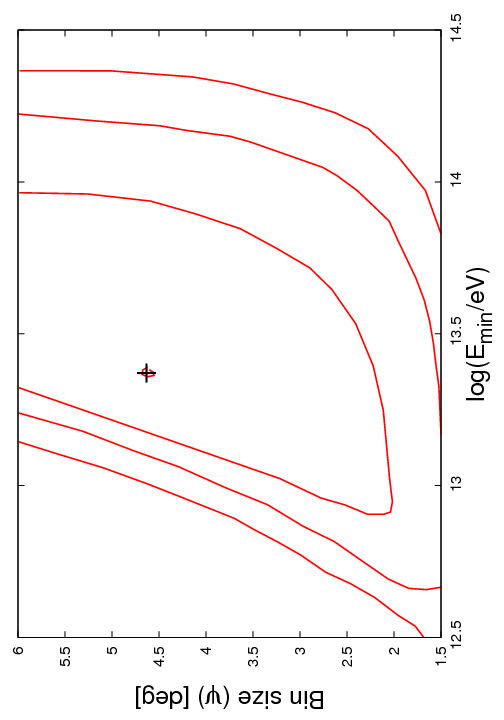}
\caption{Contour plot of the normalization factor $\Phi_0(\psi,E_{min})$ for $\gamma_{\nu}=1$ and a declination of $38^\circ$ (corresponding to Mrk 421). Contours represent the constant levels of $\Phi_0(\psi,E_{min})$: from $2.5\cdot10^{-19} [\mbox{GeV/cm}^2s]$ to $3.25\cdot10^{-19}  [\mbox{GeV/cm}^2s]$ with step of 0.25, from center to the border, for a scale factor of $E_0=10^{17}eV$. The optimal choice of $\psi,E_{min}$ corresponding to the minimum of $\Phi_0(\psi,E_{min})$ marked by the blue cross.}
\label{fig:paramspace}
\end{figure}

Fig.~\ref{fig:paramspace} shows an example of the calculation of $\Phi_0(\psi, E_{min},\gamma_{\nu})$ for a particular choice of $\gamma_{\nu}=1$ assuming the source declination $\delta=38^\circ$. From this figure one could see that for such a hard source spectrum the best choice of $E_{min}$ is $E_{min}\simeq 20$~TeV. Such a high value of $E_{min}$ efficiently suppresses the atmospheric neutrino background, so that the radius of the circle $\psi$ from which the source signal could be collected could be rather large, $\psi\simeq 4.6^\circ$.

When the spectral index increases, the spectrum of the particles distribution becomes softer and the optimal value of $E_{min}$ decreases (as expected, since there are more of particles of lower energies in the source spectrum). Simultaneously, the optimal angular size of the signal region $\psi$ is also reduced to avoid too high number of background events.  Therefore, the cross point corresponding to the optimal choice of $(\psi, E_{min})$ moves to the lower left corner of the plot in Fig. \ref{fig:paramspace}.

\begin{figure}
\includegraphics[height=\columnwidth,angle=270]{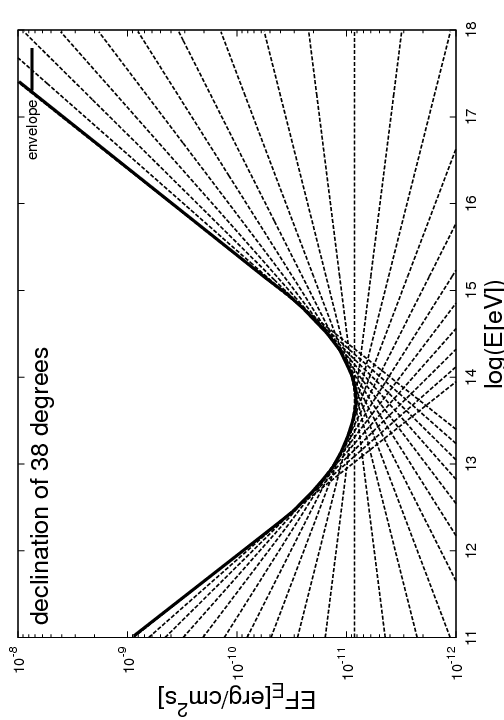}
\caption{A set of the powerlaw source spectra $\Phi_0*\left(E/E_0\right)^\gamma_{\nu}$ for different choices of $\gamma_{\nu}$ for muonic neutrinos. Black thick curve is an envelope of all the powerlaw straight lines. The declination of the source is assumed to be $\delta=38^\circ$, corresponding to the declinations of  Mrk421. }
\label{fig:envelope}
\end{figure}

We find the optimal values $\psi_*$ and $E_{min,*}$ and minimal value of the normalization $\Phi_{0*}(\gamma_{\nu},\delta)$ corresponding to $(\psi_*, E_{min,*})$. Fig. \ref{fig:envelope} shows a set of the powerlaw spectra (\ref{source}) for a range of different values of $\gamma_{\nu}$ with normalizations $\Phi_{0*}$ corresponding to the best choice of parameters $(\psi, E_{min})$.  Each straight line in this figure represents an upper limit on the neutrino flux from a source with a powerlaw spectrum for different powerlaw slopes $\gamma_{\nu}$. From this figure one could see that a convenient way to represent the upper limit for an arbitrary choice of $\gamma_{\nu}$ would be to plot an envelope curve for all the straight lines (the black thick curve). The sense of this curve is the following. For any value of $\gamma_{\nu}$, the upper limit on the source spectrum is tangent to the envelope curve.  We have verified that in the case of $\gamma_{\nu}=2$, the upper limits on the fluxes of sources at different declinations, calculated using the technique discussed above, coincide with those reported by \citet{ic40_limits} for the median sensitivity.

\end{document}